\documentclass[twocolumn,floatfix,superscriptaddress,a4paper,
               showpacs,showkeys,nofootinbib,reprint,prc]{revtex4}
\usepackage{epsfig}
\usepackage{latexsym}
\usepackage{xspace}
\usepackage{hyperref}
\usepackage[latin2]{inputenc}
\usepackage{indentfirst}
\usepackage{enumerate}
\usepackage{color}

\usepackage{amsmath}
\usepackage{amssymb}
\usepackage[english]{babel}
\usepackage{url}
\newcommand{\eq}[1]{\begin{align} #1 \end{align}}
\begin{document}

%TC:ignore

\title{Critical point of nuclear matter and beam energy dependence \\ of net proton number fluctuations}
\author{Volodymyr Vovchenko}
\affiliation{Institut f\"ur Theoretische Physik, Goethe Universit\"at Frankfurt, D-60438,
Frankfurt am Main, Germany}
\affiliation{Frankfurt Institute for Advanced Studies, Giersch Science Center,
D-60438, Frankfurt am Main, Germany}
\author{Lijia Jiang}
\affiliation{Frankfurt Institute for Advanced Studies, Giersch Science Center,
D-60438, Frankfurt am Main, Germany}
\author{Mark I. Gorenstein}
\affiliation{Frankfurt Institute for Advanced Studies, Giersch Science Center,
D-60438, Frankfurt am Main, Germany}
\affiliation{Bogolyubov Institute for Theoretical Physics, 03680 Kiev, Ukraine}
\author{Horst Stoecker}
\affiliation{Institut f\"ur Theoretische Physik, Goethe Universit\"at Frankfurt, D-60438,
Frankfurt am Main, Germany}
\affiliation{Frankfurt Institute for Advanced Studies, Giersch Science Center,
D-60438, Frankfurt am Main, Germany}
\affiliation{GSI Helmholtzzentrum f\"ur Schwerionenforschung GmbH, D-64291 Darmstadt,
Germany}

\begin{abstract}
The beam energy dependence of net baryon number susceptibilities
is studied in the framework of the hadron resonance gas model with the
attractive and repulsive van der Waals interactions between baryons.
The collision energy dependences for the skewness  $S\sigma$ and kurtosis $\kappa\sigma^2$
deviate significantly from the Poisson baseline and demonstrate the existence of rich structures at moderate collision energies.
This behavior may result from the critical end point of the nuclear liquid-gas first order phase transition.
In particular, $\kappa\sigma^2$ shows a non-monotonic energy dependence,
and, in contrast to the standard scenario for the QCD critical point,
it does not decrease at low collision energies.
It is also found that the measurable net proton fluctuations differ significantly from the net baryon fluctuations when interactions between baryons cannot be neglected.
The results are compared with
the experimental net proton number fluctuations measured by the STAR collaboration.
\end{abstract}


\maketitle

%TC:endignore

\section{Introduction}

One of the main goals of today's  experiments in nucleus-nucleus (A+A)
collisions is to search for
the critical point (CP) of QCD matter
\cite{Stephanov:1998dy,Stephanov:1999zu,Stephanov:2004wx,Lacey:2006bc,Aggarwal:2010cw}
(see also recent reviews
 \cite{Luo:2017faz, Gazdzicki:2015ska}).
Theoretical arguments suggest the enhancement of
net baryon number fluctuations in the critical region
\cite{Stephanov:1998dy,Stephanov:1999zu,Athanasiou:2010kw,Hatta:2003wn,Stephanov:2008qz,Kitazawa:2012at}.
On
the experimental side, the STAR collaboration has presented the Beam Energy Scan
data of proton cumulants in Au+Au collisions for center
of mass energies per nucleon pair, $\sqrt{s_{NN}}$,
of the Relativistic Heavy
Ion Collider
from $7.7$~GeV to $200$~GeV. At moderate collision
energies, the data \cite{Aggarwal:2010wy,Adamczyk:2013dal,Luo:2015ewa} of
skewness $S\sigma$ and kurtosis $\kappa\sigma^2$ of the net proton number
fluctuations  show an interesting non-monotonic
behavior and exhibit large
deviations from the Poisson baseline.
This is considered as a possible signal for the CP
\cite{Jiang:2015hri,Ling:2015yau,Mukherjee:2015swa,Mukherjee:2016kyu,Herold:2016uvv,Bluhm:2016byc,Jiang:2017mji}.
The experimental data
is also influenced by other effects,
such as initial state fluctuations~\cite{Kapusta:2011gt},
system volume fluctuations~\cite{Gorenstein:2011vq,Sangaline:2015bma,Xu:2016skm}, stopping effects~\cite{Bzdak:2016jxo}, acceptance effects~\cite{Bzdak:2012ab,Braun-Munzinger:2016yjz},
global charge conservation~\cite{Begun:2004gs,Bzdak:2012an},
effects of the hadronic phase~\cite{Kitazawa:2012at,Steinheimer:2016cir}, etc.
Some of these effects have been studied with transport models~\cite{Xu:2016qjd,He:2016uei,He:2017zpg}.
It has also been argued that correlation functions, expressible through cumulants, may provide a cleaner information about the underlying dynamics in heavy-ion collisions~\cite{Bzdak:2016sxg}.

In this paper, we focus on the interactions between baryons
within the Quantum van der Waals (QvdW) equation of state \cite%
{Vovchenko:2015xja,Vovchenko:2015vxa,Vovchenko:2015pya}.
In Ref.~\cite{Vovchenko:2016rkn}
the Hadron Resonance Gas (HRG) model with QvdW   interactions
between baryons and between antibaryons was formulated and compared
with lattice QCD simulations at zero chemical potentials
 in the crossover temperature region. Inclusion of the
QvdW interactions has only minor influence on the pressure and energy density
in comparison with the ideal HRG~(IHRG) model,
but they change significantly the structure of all high order fluctuations
of conserved charges, in most cases leading to a much better agreement with the lattice data,
e.g., a quantitative agreement up to $T \simeq 160$~MeV is obtained for
the net baryon number susceptibilities.
The influence of nucleon-nucleon interactions and the associated nuclear liquid-gas criticality on baryon number fluctuations had previously also been pointed out in Refs.~\cite{Fukushima:2014lfa,Mukherjee:2016nhb}.
In the
following, we calculate the baryon number fluctuations
along the chemical freeze-out line within the QvdW-HRG model and
make a comparison with the STAR data.
We employ the QvdW formalism in this paper because it is the simplest and straightforward way
to include the nuclear matter physics into the hadronic equation of state.

\section{Model}

Searching for the CP signatures~(or, more generally, for any notable deviations from the Poisson baseline) within an equilibrium HRG model is only possible if appropriate hadron-hadron interactions are taken into account.
Following \cite{Vovchenko:2016rkn}
we assume that
the pressure of the QvdW-HRG system can be written as the sum of $3$ terms:
\begin{equation}
p\left( T,\mu \right) =p_{M}\left( T,\mu \right) +p_{B}\left( T,\mu \right)
+p_{\bar{B}}\left( T,\mu \right)~ ,\label{p}
\end{equation}%
where partial pressures of mesons, baryons, and antibaryons are given by
\begin{eqnarray}
p_{M}\left( T,\mu \right) &=&\sum_{j\in M}p_{j}^{id}\left( T,\mu _{j}\right)~
, \label{pM}\\
p_{B(\bar{B})}\left( T,\mu \right)& =& \sum_{j\in B(\bar{B})}
p_{j}^{id}\left( T,\mu _{j}^{B(\bar{B}) \ast }\right)
-an_{B(\bar{B}) }^{2}. \label{p-BantiB}
\end{eqnarray}%
Here
$p_{j}^{\rm id}$ are the ideal Bose-Einstein~[Eq.~(\ref{pM})] or Fermi-Dirac~[Eq.~(\ref{p-BantiB})] pressures,
$ \mu=(\mu_B,\mu_S,\mu_Q)$
are the chemical potentials which regulate the average values of the total baryonic number $B$,
strangeness $S$, and electric charge $Q$.
In Eqs.~(\ref{pM})-(\ref{p-BantiB}),
$\mu_j=b_j\mu_B+s_j\mu_S+q_j\mu_Q$ and
\eq{\label{muB*}
\mu _j^{B*}
& = \mu_j - b\,p_{B}
- a\,b\,n_{B
}^2 + 2\,a\,n_{B
}~,\\
n_{B} & =\sum_{j \in B}n_j =
\left( 1-bn_B\right) \sum_{j\in
B} n_{j}^{id}\left( T,\mu_j^{B*}\right)~,\label{nB}
}
are, respectively, the shifted chemical potentials for different baryons
and total density of all baryons.
Note that Eq.~\eqref{p-BantiB} implicitly contains terms of arbitrary large powers of $n_{B(\bar{B})}$, through the transcendental equations~\eqref{muB*} and \eqref{nB}.
The expressions for $\mu_j^{\bar{B}*}$ and $n_{\bar{B}}$ for the antibaryons are analogous
to (\ref{muB*}) and (\ref{nB}).
The net baryonic density is $\rho_B=n_B-n_{\bar{B}}$.
The QvdW interactions are assumed to exist between all pairs of baryons
and between all pairs of antibaryons,
including all the strange (anti)baryons,
with the same parameters
as for nucleons, $\ a=329$ MeV fm$^{3}$ and $b=3.42$ fm$^{3}$~\cite{Vovchenko:2016rkn}. These $a$ and $b$ parameters
were obtained in Ref.~\cite{Vovchenko:2015vxa} by fitting the
saturation density, $n_{0}=0.16$ fm$^{-3}$, and binding energy, $E/A=-16$ MeV,
of the ground state of nuclear matter.
Baryon-antibaryon, meson-meson, and
meson-(anti)baryon QvdW interactions are neglected. Thus, the present version
of the QvdW-HRG model is a "minimal interaction" extension of the IHRG model.
It should be noted that the vdW parameters $a$ and $b$ may attain different values in the meson-dominated
region of the phase diagram at small $\mu_B / T$ and high $T$
(see, e.g.,
the recent analysis of the lattice data~\cite{Vovchenko:2017drx}).
As our focus here is on the effects of the nuclear liquid-gas criticality in the baryon-rich region,
we retain the $a$ and $b$ parameter values found from nuclear ground state properties.
For the same reason we omit the possible vdW-like interactions between baryons and antibaryons.
Since we do not study strangeness related observables,  we do not consider also
the possible
inclusion
of the uncharted strange baryons suggested in~\cite{Alba:2017mqu}.
The sums in Eqs.~(\ref{pM}) and (\ref{p-BantiB}) include
all stable hadrons and resonances listed in the Particle Data Tables~\cite{Patrignani:2016xqp}.

The QvdW-HRG equation of state (\ref{p})-(\ref{nB}) leads to the liquid-gas phase transition
in the symmetric nuclear matter, with a CP at
$T_c \cong 19.7$~MeV and
$\mu^c _{B} \cong 908$~MeV \cite{Vovchenko:2015vxa}, where a singular behavior of baryon number fluctuations appears~\cite{Vovchenko:2015pya}.

\begin{figure*}[tb]
\center
\includegraphics[width=0.49\textwidth]{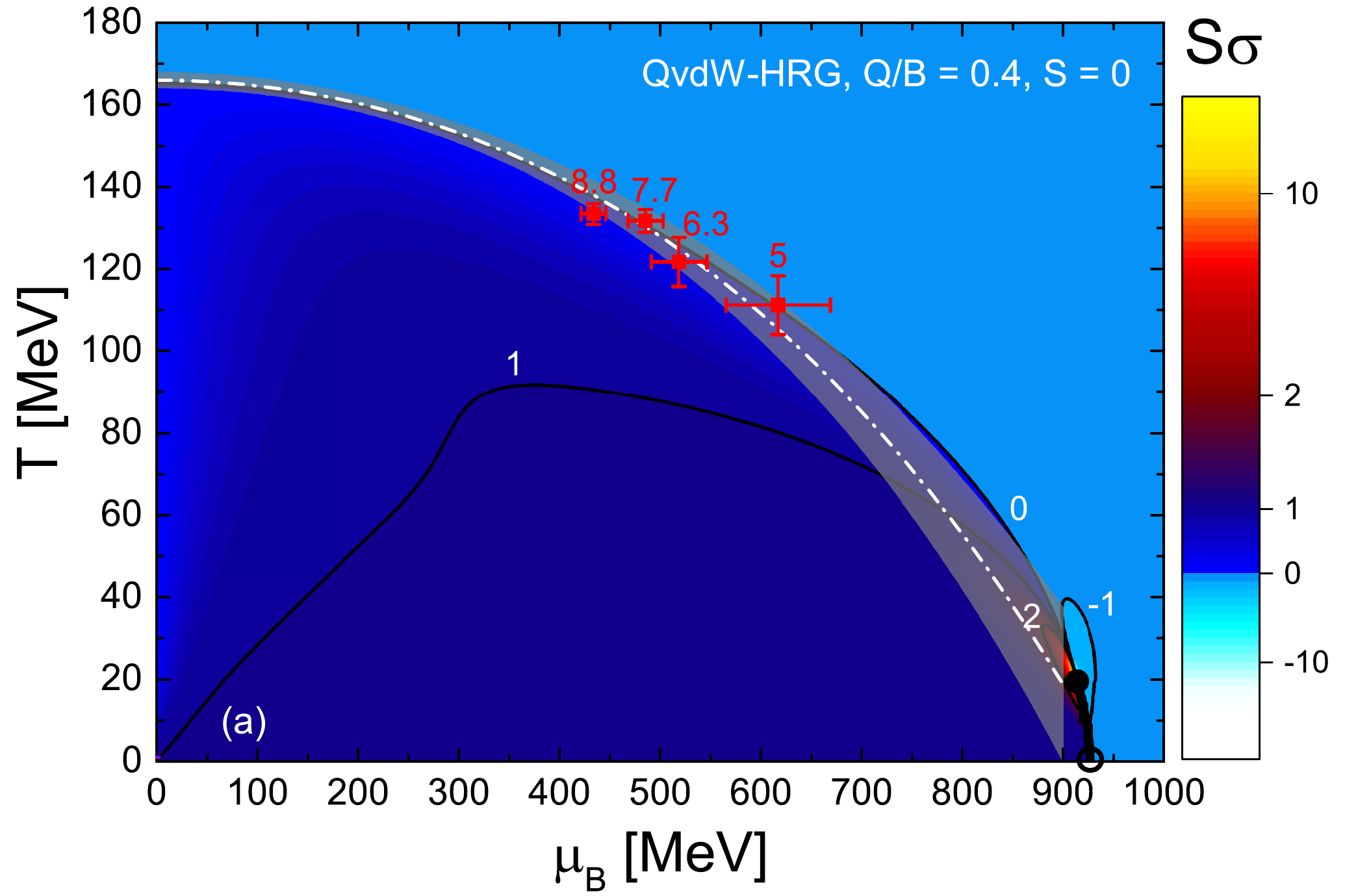}
\includegraphics[width=0.49\textwidth]{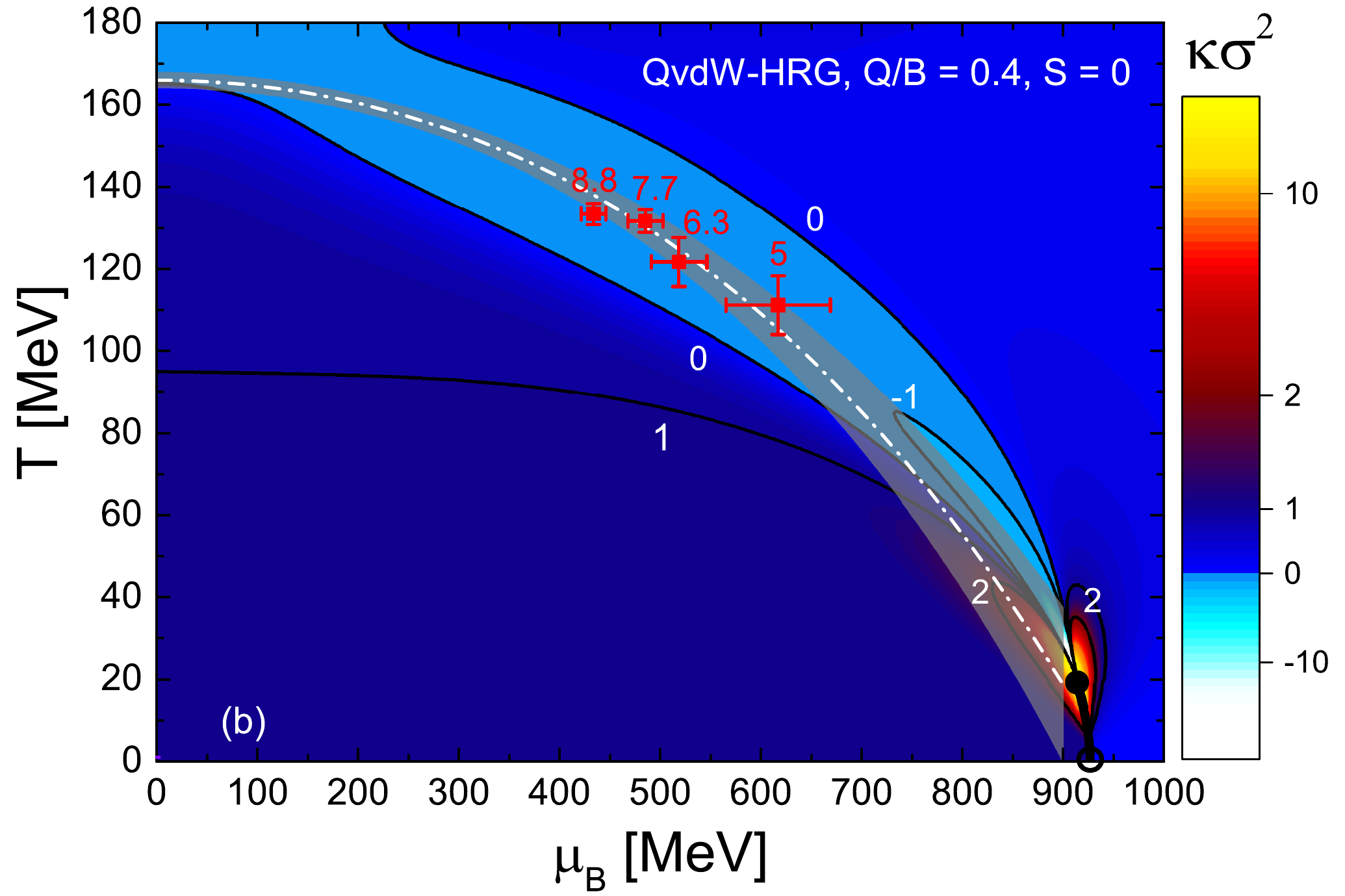}
\caption{The contour plots of (a) $S\protect\sigma$ and (b) $\protect\kappa\protect\sigma^2$ for net baryon fluctuations in the $\mu_B$-$T$ plane, as calculated within the QvdW-HRG model.
The dash-dotted line shows the IHRG model chemical freeze-out curve~[Eq.~\ref{Tmu}] from Ref.~\cite{Cleymans:2005xv},
while the semi-transparent shaded area along it depicts the uncertainty in the parameters of this freeze-out curve. The nuclear liquid-gas phase transition is depicted by the thick black line which ends at the CP depicted by full circle.
The red circles with error bars correspond to the thermal fits performed within
the QvdW-HRG model to the hadron yield data at AGS~($\sqrt{s_{NN}} = 5$~GeV)
and SPS~($\sqrt{s_{NN}} = 6.3$, 7.7, and 8.8~GeV).
}
\label{figTmu}
\end{figure*}

To calculate the particle number fluctuations in A+A collisions
we adopt the thermodynamic freeze-out parameters, which were obtained in Refs.~\cite{Andronic:2005yp,Cleymans:2005xv}
by fitting the particle yields at different collision
energies within the HRG model.
The following simple functional form of the freeze-out curve was obtained~\cite{Cleymans:2005xv}:
\eq{
T =a_{1}-a_{2}\mu_B^{2}-a_{3}\mu_B^{4}~,~~~~
\mu_B=\frac{b_{1}}{1+b_{2}\sqrt{s_{NN}}}, \label{Tmu}
}
with $a_{1}=0.166 \pm 0.002$ GeV, $a_{2}=0.139 \pm 0.016$ GeV$^{-1}$, $a_{3}=0.053 \pm 0.021$ GeV$^{-3}$, $b_{1}=1.308 \pm 0.028$ GeV, $b_{2}=0.273 \pm 0.008$ GeV$^{-1}$.
A flatter chemical freeze-out curve,
with a possibly
lower limiting $T \sim 145$ MeV value at $\mu_B = 0$, was also suggested
in Refs.~\cite{Alba:2014eba,Vovchenko:2015idt,Bazavov:2017dus,Critelli:2017oub} based on fluctuations of conserved charges.
As our focus here is
the
baryon-rich region, we retain the original parameterization of Ref.~\cite{Cleymans:2005xv}.
It was obtained by analyzing the hadron yield data at collision energies as low as Schwerionen Synchrotron (i.e., $\mu_B / T \simeq 15$).

The
fluctuations of conserved charges
can be calculated in the grand canonical ensemble (GCE) from the system pressure
by taking the derivatives over the corresponding chemical potentials.
The net baryon number fluctuations are given by the following  normalized cumulants
(susceptibilities) $(n=1,\ldots,4)$:
\eq{
\chi_{n}& =\frac{\partial ^{n}\left( p/T^{4}\right) }{\partial \left( \mu_B
/T\right) ^{n}}= \frac{\partial ^{n}\left( p_B/T^{4}\right) }{\partial \left( \mu_B
/T\right) ^{n}}\nonumber \\
&+~(-1)^n\frac{\partial ^{n}\left( p_{\bar{B}}/T^{4}\right) }{\partial \left( \mu_{\bar{B}}
/T\right) ^{n}}~\equiv~\chi_n^{B}+(-1)^n \chi_n^{\bar{B}}~,\label{kn}
}
where $\mu_{\bar{B}} \equiv - \mu_B$
is the baryochemical potential~(not to be confused with the shifted chemical potentials $\mu^{B*}_j$ which are only auxiliary quantities).
The simple presentation of the $\chi_n$, with the $\chi_n^B$ and $\chi_n^{\bar{B}}$ cumulants in (\ref{kn})
is due to the
absence of correlations between  baryons and antibaryons in the QvdW-HRG model, i.e.
the probabilty distribution ${\cal P} (N_B,N_{\bar{B}})$ of the number of baryons
and antibaryons is the product
 $ {\cal P} (N_B,N_{\bar{B}})=
{\cal P}_B(N_B)\,{\cal P}_{\bar{B}}(N_{\bar{B}})$.

In the following, we consider the normalized skewness and kurtosis for the net baryonic number
fluctuations. They are defined as the corresponding ratios of cumulants:
\begin{eqnarray}
S\sigma  =\frac{\chi_{3}}{\chi_{2}}~,~~~~
\kappa \sigma ^{2} =\frac{\chi_{4}}{\chi_{2}}~.
\label{kurt}
\end{eqnarray}

\section{Results and discussion}

The $\mu_B$-$T$ contour plots of $S \sigma$ and $\kappa \sigma^2$, as calculated in the QvdW-HRG model, are depicted in Fig.~\ref{figTmu}.
The IHRG model chemical freeze-out curve (\ref{Tmu}) from Ref.~\cite{Cleymans:2005xv}
is depicted in Fig.~\ref{figTmu} by the dash-dotted line.
At each $\mu_B$-$T$ point, the strangeness and electric charge chemical potentials $\mu_S$ and $\mu_Q$ are determined in the QvdW-HRG model from the condition of strangeness neutrality, $S = 0$, and fixed electric-to-baryon charge ratio, $Q / B = 0.4$.
Figure~\ref{figTmu} shows that signals from the nuclear matter CP shine brightly in net baryon $S \sigma$ and $\kappa \sigma^2$ across the whole phase diagram probed by the heavy-ion collision experiments.

\begin{figure*}[tb]
\center
\includegraphics[width=0.49\textwidth]{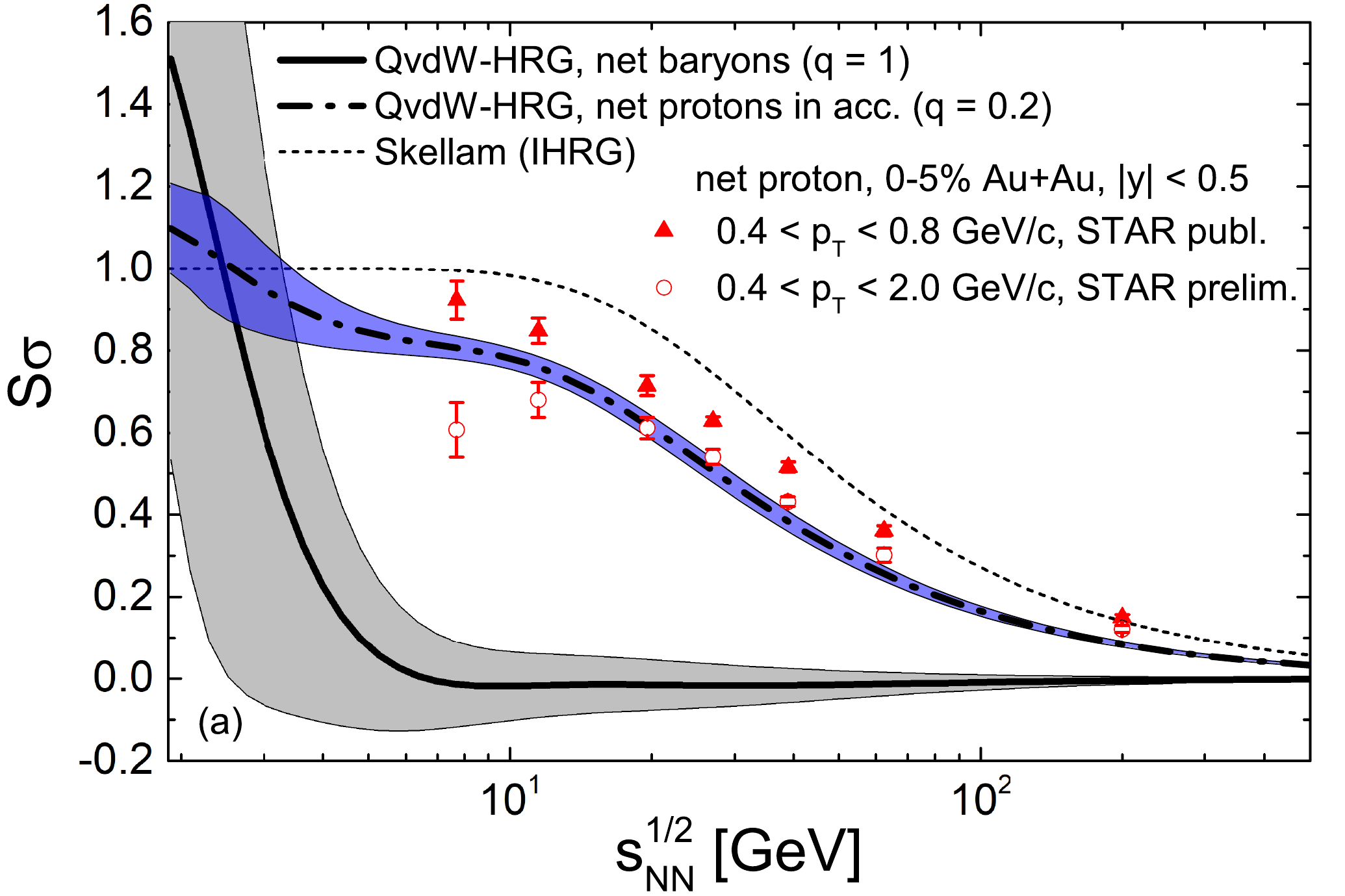}
\includegraphics[width=0.49\textwidth]{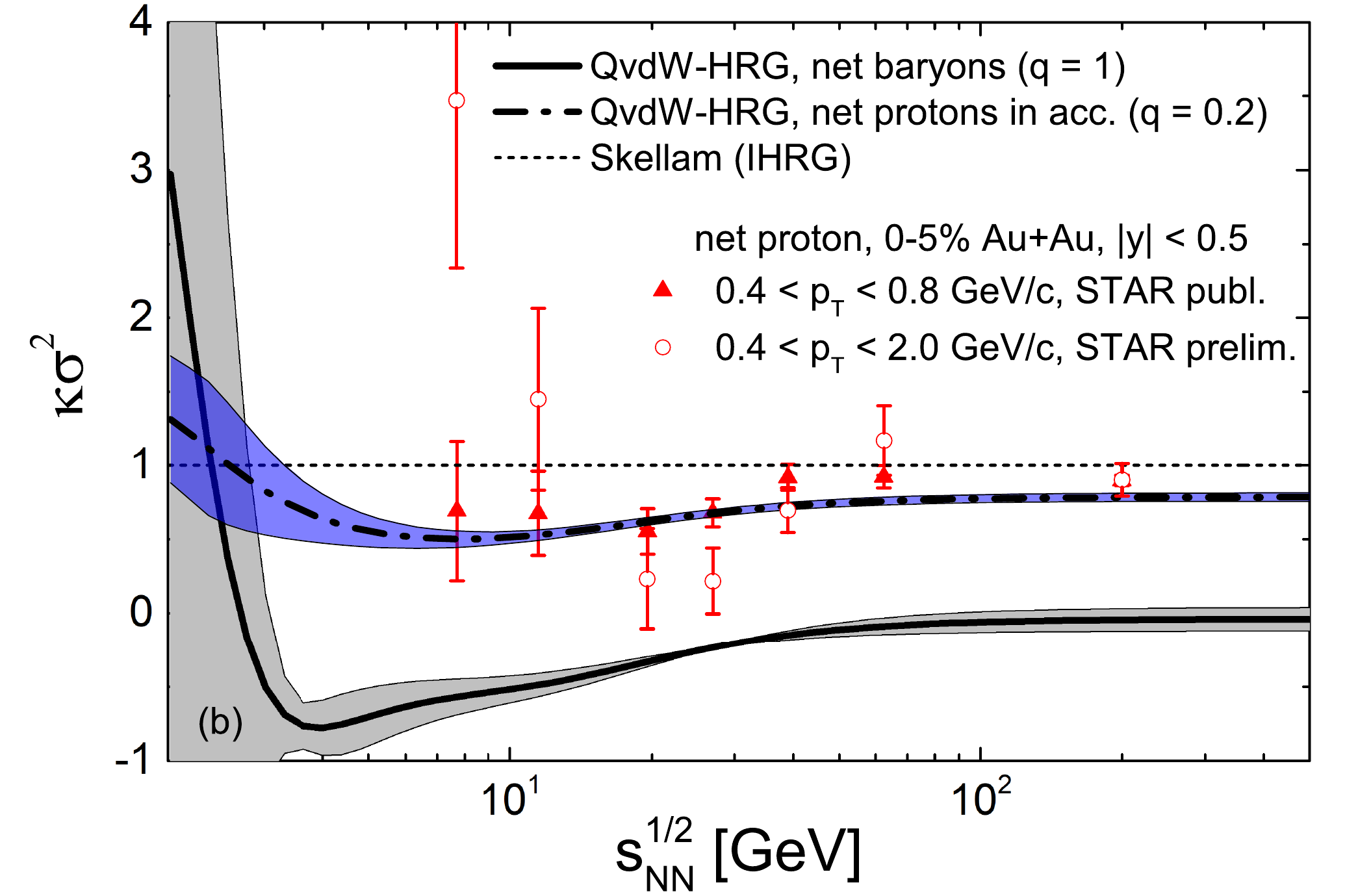}
\caption{The (a) $S\protect\sigma$ and (b) $\protect\kappa\protect\sigma^2$
of net baryons in full acceptance (solid lines) and
net protons in finite acceptance (dash-dotted lines)
as a function of the collision energies in the QvdW-HRG model.
The bands estimate the uncertainty coming from the chemical freeze-out curve~[Eq.~\eqref{Tmu}].
The dotted lines correspond to the Skellam distribution baseline of non-interacting hadrons.
The STAR collaboration data for the midrapidity net proton fluctuations in the $0.4 < p_T < 0.8$~GeV/$c$~\cite{Adamczyk:2013dal} and $0.4 < p_T < 2$~GeV/$c$~\cite{Luo:2015ewa} intervals are shown by full and open red circles, respectively.
}
\label{fig1}
\end{figure*}

The mean values, $\langle N_B\rangle$, and
central moments,
$m_n^B=\sum_{N_B}(N_B-\langle N_B\rangle )^n {\cal P}_B(N_B)$,
of the corresponding baryon number
distributions may be measured in A+A collisions.
From these measured values the cumulants of ${\cal P}(N_B)$ distribution are found:
\eq{\label{cum}
& K_1^B=\langle N_B\rangle, ~~~~K_2^B=m_2^B~,~~~~K_3^B=m_3^B~,\nonumber \\
& K_4^B=m_4^B-3m_2^B~.
}
Similar expressions hold for antibaryon quantities.

The cumulants $K_n$  are
connected to $\chi_n$ in Eq.~(\ref{kn}) as $K_n=VT^3\,\chi_n$, where $V$ is the system volume.
Therefore, all ratios of cumulants $\chi_n$, in particular those in Eq.~(\ref{kurt}),
are equal to the corresponding ratios of $K_n$.
The
"required acceptance"
\cite{Jeon:2000wg,Asakawa:2000wh,Bzdak:2012ab} for the event-by-event measurements in A+A
reactions should then satisfy the following requirements:
The
GCE
can be used
if only the accepted phase-space region is a {\it small} part of the whole system.
On the other hand, this region  should be {\it large} enough to capture
the relevant physics.

Following Refs.~\cite{Kitazawa:2012at,Bzdak:2012ab}, it is assumed
that acceptance corrections from all different sources can be
modeled by binomial distributions,
\eq{\label{binom}
P\left( n \right) =\sum_{N=n}^{\infty }{\cal P}\left( N\right)
\frac{N!}{n!\left( N-n\right)!} q^{n} \left( 1-q\right) ^{N-n},
}
where $n$ represents the \emph{measured} number of baryons (or antibaryons), and $N$ represents
their \emph{true} numbers.
Equation~\eqref{binom}  includes the possible effects of isospin randomization, 
which are also modeled by the binomial distribution~\cite{Kitazawa:2011wh,Kitazawa:2012at}.
The parameter $q$~($0\le q\le 1$) describes the acceptance effects.
In general, it can be different
for baryons and antibaryons.
Equation~(\ref{binom}) gives $P(n)={\cal P}(N)$ for $q=1$, whereas
$P(n)$ becomes the Poisson distribution, with $\langle n\rangle=q\langle N\rangle$ in the limit $q\rightarrow 0$.
Using Eq.~(\ref{binom}), the calculation of all moments and all cumulants $c_n$, for all
accepted baryons and antibaryons,
are straightforward.
They are presented as linear combinations,
$c_n=a_1K_1+\ldots+a_nK_n$, with $a_i=a_i(q)$ (see details
and explicit expressions in Refs.~\cite{Bzdak:2012ab,Kitazawa:2012at}).
The STAR data correspond to accepted, efficiency corrected protons and antiprotons at midrapidity~($|y|<0.5$),
in two different transverse momentum intervals:
$0.4 < p_T < 0.8$~GeV/$c$~\cite{Adamczyk:2013dal}
and $0.4 < p_T < 2$~GeV/$c$~\cite{Luo:2015ewa}.
It is further assumed that parameters $q$ for baryons and antibaryons attain the same value.
We take $q=1$ and $0.2$,
which approximately represents
two cases:
1) the ideal case of all baryons and antibaryons~($q = 1$);
2) a more realistic case of protons and antiprotons within a particular acceptance~($q = 0.2$),
including the possible isospin randomization~\cite{Kitazawa:2011wh,Kitazawa:2012at}.

Fig.~\ref{fig1} shows the skewness
and the kurtosis, as calculated in the QvdW-HRG
model, as functions of $\sqrt{s_{NN}}$
along the chemical freeze-out line (\ref{Tmu}). Solid lines represent the results
of $S\sigma$ and $\kappa\sigma^2$ of the QvdW-HRG model under the assumption of a full acceptance
for both baryons and antibaryons, $q=1$ in Eq.~(\ref{binom}).
The kurtosis $\kappa\sigma^2$ shows nonmonotonic
behavior at moderate collision energies,
similar to the STAR data
for net protons \cite{Luo:2015ewa}.
Dashed-dotted lines
represent the results for $q=0.2$.
Obviously, acceptance effects have a large quantitative and qualitative
influence on the behavior of both $S\sigma$ and $\kappa\sigma^2$,
and appear to bring them closer to the experimental measurements.
Note that the model does not reproduce the preliminary STAR data at the lowest collision energies.
This could signal new physics not contained in the purely hadronic QvdW-HRG model, although it may be more prudent to await for these data to be finalized before stronger conclusions can be drawn.

At small $q$, the QvdW-HRG results approach the baselines
obtained from ideal gas calculations, shown in Fig.~\ref{fig1} by the dotted lines.
Note that there would be no difference between net baryon and accepted
net proton cumulant ratios in the IHRG model.
This is because the binomial filter acts as a ``Poissonizer'', and therefore it
does not introduce differences between cumulant ratios of net proton and net baryon distributions in the IHRG model,
where fluctuations correspond to Poisson statistics.
More detailed IHRG model studies~\cite{Nahrgang:2014fza} show that net proton fluctuations remain very 
similar to net baryon fluctuations in the IHRG model even when more effects, such as the probabilistic resonance decays, are taken into account.
In contrast to the IHRG model,
the presence of baryon-baryon interactions in the QvdW-HRG model makes the net baryon distribution quite different from the Poisson statistics.
In this case, the application of the binomial filter changes the cumulant ratios and is the reason for
the large difference between the results for net baryon~($q = 1$) and
accepted net proton~($q = 0.2$) fluctuation observables seen in Fig.~\ref{fig1}.

The binomial filter is only a schematic way to do an acceptance correction, and a more accurate analysis should take into account the correlation range relative to the acceptance.
Nevertheless, the binomial filter is fully sufficient to illustrate 
that the presence of the QvdW interactions between baryons at the chemical freeze-out leads to differences between experimentally observed net proton fluctuations and net baryon fluctuations.
This difference can be quite large when baryon-baryon interactions are non-negligible, as suggested by our calculations within the QvdW-HRG model.
If the effects of baryonic interactions are indeed significant, then the justification for the direct correspondence between the net
baryon cumulant ratios, calculated either from first principles
in lattice QCD~\cite{Borsanyi:2014ewa,Bazavov:2017tot} or within
effective models for QCD equation of state~\cite{Albright:2015uua,Fu:2016tey,Almasi:2017bhq},
and the net proton cumulant ratios measured in heavy-ion collisions at RHIC, can be questioned.
Corrections for differences between net proton and net baryon fluctuations are then required.

We also stress an importance of including the full
spectrum of baryonic resonances, which is a new element compared to the earlier works~\cite{Fukushima:2014lfa,Vovchenko:2015pya}.
The resonance decay feeddown to the final proton yield could be neglected in the very vicinity of the nuclear liquid-gas transition~($T \lesssim 30$~MeV),
but it is essential at the higher temperatures probed by heavy-ion collisions: the resonance decay feeddown accounts for about 10\% of all observed protons already at the HADES energy of $\sqrt{s_{NN}} = 2.4$~GeV, and for about 50\% of all observed protons at the lowest STAR-BES energy of $\sqrt{s_{NN}} = 7.7$~GeV.
The feeddown from unstable mass fragments can also be important~\cite{Hahn:1986mb}.
In the present work we took into account the resonance decay contribution approximately, by applying the binomial filter to all baryons and antibaryons.
The full probabilistic decay treatment was considered in Ref.~\cite{Nahrgang:2014fza} for the IHRG model, a consistent result with binomial filter was reported: no additional significant differences between net proton and net baryon cumulant ratios.
It will be interesting to consider the full probabilistic 
decay treatment in the QvdW-HRG model as well to verify the accuracy of the binomial filter.

One more comment is appropriate here: The chemical
freeze-out  line (\ref{Tmu}) used in our studies is determined from the thermal fits to heavy-ion hadron yield data within the simple statistical model for non-interacting hadrons -- the IHRG model.
This may be approximately valid
in the QvdW-HRG model at large collision energies.
However, as thermal fits are affected by hadronic interactions~\cite{Vovchenko:2015cbk,Vovchenko:2016ebv},
it is not clear whether a simple IHRG model
is appropriate for determination of the chemical freeze-out conditions in the baryon-rich matter created in heavy-ion collisions at $\sqrt{s_{NN}}=7.7$~GeV and at lower collision energies.
As a cross-check, we have performed thermal fits to the hadron yield data
in central heavy ion collisions at AGS~($\sqrt{s_{NN}} = 5$~GeV)~\cite{AGSdata} and SPS~($\sqrt{s_{NN}} = 6.3$, 7.7, and 8.8~GeV)~\cite{NA49data}
within the chemical equilibrium QvdW-HRG model.
The results are depicted in Fig.~\ref{fig1} by the red symbols.
This leads to the increased uncertainties of
the $T$ and $\mu_B$ chemical freeze-out values.
Nevertheless, the overall picture is consistent with the chemical freeze-out curve given by Eq.~\eqref{Tmu}.
In fact,
in the IHRG model itself the chemical freeze-out parameters
are also rather uncertain, as illustrated
in Fig.~\ref{figTmu}.
Further refinements will be studied in the ongoing and  future heavy-ion experiments, such as the HADES experiment~\cite{Agakishiev:2009am},
the NA61/SHINE experiment~\cite{Abgrall:2014xwa,Gazdzicki:2015ska}
the STAR fixed-target program~\cite{Meehan:2016iyt},
the future CBM experiment at FAIR~\cite{Ablyazimov:2017guv}, and the future NICA project~\cite{Kekelidze:2016wkp}.
The
{\it moderate} collision energies, $\sqrt{s_{NN}} \lesssim 7.7$~GeV, look as the most interesting region
for the studies of baryon number fluctuations.

We emphasize that our results give a {\it qualitative}
description of the net proton fluctuations measured at mid-rapidity in heavy-ion collision experiments.
A complete  analysis has to take into account other effects
including the initial state fluctuations and global charge conservation,
as well as possible loss of information about equilibrium fluctuations during the non-equilibrium evolution in the hadronic phase~\cite{Steinheimer:2016cir}.
A dynamical model, incorporating the above effects,
and the baryonic interactions, can be used in a {\it quantitative} study.

\begin{figure}[tb]
\center
\includegraphics[width=0.49\textwidth]{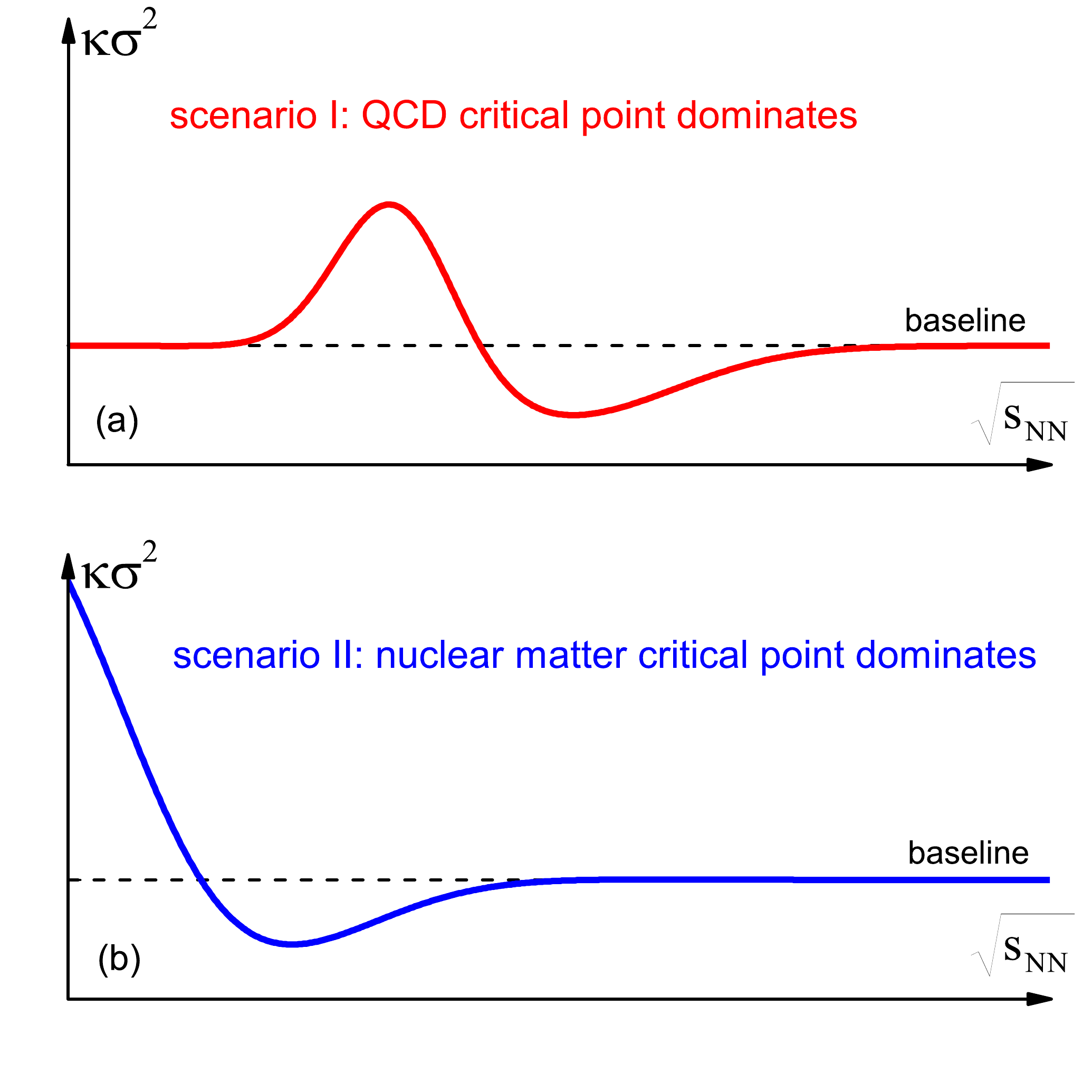}
\caption{
A schematic view of the collision energy dependence of kurtosis $\kappa \sigma^2$ of net proton fluctuations in two different scenarios.
(a): The ``standard scenario''~\cite{Stephanov:2011zz}, where the energy dependence is determined by the chiral CP of QCD.
(b): The scenario where the $\kappa \sigma^2$ behavior is determined by the nuclear liquid-gas criticality.
}
\label{figscen}
\end{figure}

\section{Summary}

In summary, the QvdW-HRG model, with attractive and repulsive vdW
interactions between the pairs of baryons and antibaryons to study the
higher order cumulants of particle number fluctuations. The cumulant ratios
which define the skewness  $S\sigma$ and the kurtosis $\kappa\sigma^2$
for baryonic number fluctuations are calculated. These quantities
show non-monotonic structures
along the chemical freeze-out line,
which bear similarities to the data presented by the STAR Collaboration.
These results emphasize the importance of the interactions between baryons for higher order fluctuations.
Any serious thermodynamics-based analysis of the net baryon fluctuation measurements should take into account the effects arising from the nuclear liquid-gas criticality.

The QvdW-HRG model predictions for higher-order net baryon fluctuations presented here are quantitatively reliable in the vicinity of the critical point of nuclear matter.
A more precise description away from nuclear matter will  require refinements and modifications.
The present results, however, are sufficient to make an important point regarding the beam energy dependence of the baryon number fluctuations.
In the "standard scenario" for the QCD CP, shown in upper panel of Fig.~\ref{figscen}, it is expected that the kurtosis $\kappa\sigma^2$ {\it decreases} with decreasing
$\sqrt{s_{NN}}$ at moderate collision energies, because the chemical freeze-out
$(T,\mu_B)$-point moves away from the hypothetical QCD CP~\cite{Stephanov:2011zz}.
In contrast, here~(lower panel of Fig.~\ref{figscen}) $\kappa\sigma^2$ keeps {\it increasing}
with decreasing $\sqrt{s_{NN}}$, as the chemical freeze-out point moves closer towards the nuclear CP.
Future fluctuation measurements and their analysis at moderate collision energies should be able to distinguish these scenarios.

%TC:ignore

\begin{acknowledgments}

%\section*{Acknowledgements}
\emph{Acknowledgments.} We are grateful to Adam Bzdak, Volker Koch, Misha Stephanov,
Jan Steinheimer,
and Nu Xu for stimulating discussions.
The authors appreciate interesting discussions with the participants at the EMMI Workshop on Critical
Fluctuations
Near the QCD Phase Boundary in Relativistic Nuclear Collisions, 10-17 October
2017, Wuhan, China.
This work was supported by HIC for FAIR within the LOEWE program of the State of Hesse.
V.V. acknowledges the support from HGS-HIRe for FAIR.
H.St. acknowledges the support through the Judah M. Eisenberg Laureatus Chair at Goethe University.
The work of M.I.G. was supported
by the Program of Fundamental Research of the Department of
Physics and Astronomy of National Academy of Sciences of Ukraine.

\end{acknowledgments}

%TC:endignore

\end{document}